\documentclass[aps,pb,amsmath,amssymb,twocolumn,twoside,showpacs,superscriptaddress,longbibliography]{revtex4-2}
\usepackage[pdftex]{graphicx}
\usepackage[colorlinks=true,citecolor=blue]{hyperref}
\usepackage{graphicx,textcomp,dcolumn,hyperref,upgreek}

\usepackage{physics}

\usepackage{url}

\usepackage{dsfont}

\usepackage{blindtext}

\newcommand{\z}{_{\noindent z}}

\begin{document}

\title{
Full counting statistics in a Majorana single-charge transistor
}
\author{Eric~Kleinherbers}
\email{eric.kleinherbers@uni-due.de}
\affiliation{Faculty of Physics and CENIDE, University of Duisburg-Essen, 47057 Duisburg, Germany}
\affiliation{Department of Physics and Astronomy, University of California, Los Angeles, California 90095, USA}

\author{Alexander~Sch{\"u}nemann}
\affiliation{Faculty of Physics and CENIDE, University of Duisburg-Essen, 47057 Duisburg, Germany}

\author{J{\"u}rgen~K{\"o}nig}
\affiliation{Faculty of Physics and CENIDE, University of Duisburg-Essen, 47057 Duisburg, Germany}
               
\date{\today}

\begin{abstract}
We study full counting statistics of electron transport through a Majorana single-charge transistor.
At low bias voltage, transport is dominated by the so-called Josephson-Majorana cycle, a sequence of normal and anomalous single-charge and Josephson tunneling. 
Factorial cumulants characterizing the full counting statistics elucidate the correlated nature of the charge transfers in this cycle. 
Moreover, we predict a topological transition in the full counting statistics from a perfect Poissonian transfer of Cooper pairs to a correlated switching between two distinct fermion parity states with increasing Josephson coupling.  
\end{abstract}
\maketitle

%%%%%%%%%%%%%%%%%%%%%%%%%%%%%%%%%%%%%%%%%%%%%%%%%%%%%%%%%%%%%%%%%%%%%%%%%%%%%%%%%%%%%%%%%%%%%%%%%%%%
%%%%%%%%%%%%%%%%%%%%%%%%%%%%%%%%%%%%%%%%%%%%%%%%%%%%%%%%%%%%%%%%%%%%%%%%%%%%%%%%%%%%%%%%%%%%%%%%%%%%
\section{\label{sec:Introduction} Introduction}
Majorana bound states are exotic, charge-neutral quasiparticle excitations in superconductors with a non-trivial topology~\cite{alicea_2012, beenaker_2013}. In recent years, they have attracted significant attention in the field of condensed matter physics due to their potential use in quantum computers~\cite{aasen_2016}.  
Being non-Abelian anyons, they fulfill braiding statistics~\cite{clarke_2017} as well as fusion rules~\cite{ruben_2022} which makes them interesting candidates for topological quantum computing~\cite{nayak_2008,sarma_2015}.

A famous model system that allows for the existence of Majorana bound states is the Kitaev chain~\cite{Kitaev_2001}, a one-dimensional topological superconducting wire. 
There, the two Majorana bound states at the ends of the wire comprise a single nonlocal fermion level at zero energy that is robust against local disturbances, a phenomenon called topological protection.
These zero-energy quasiparticles give rise to interesting transport properties, such as electron teleportation~\cite{fu_2010}, a zero-bias peak~\cite{nichele_2017}, a $4\pi$-Josephson effect~\cite{laroche_2019} or the parity blockade~\cite{nitsch_2022}.
Furthermore, the combination of multiple Majorana states~\cite{ekstroem_2020} as well as the effect of the Majorana states on the charge dynamics in current-biased Josephson junctions~\cite{vanbeek_2018} have been considered.
Essential features of Majorana states are already contained in interacting double quantum dots~\cite{leijnse_2012,tsintzis_2022}.

\begin{figure}[h]
\vspace{1.cm}		
	\includegraphics[width=8.6cm]{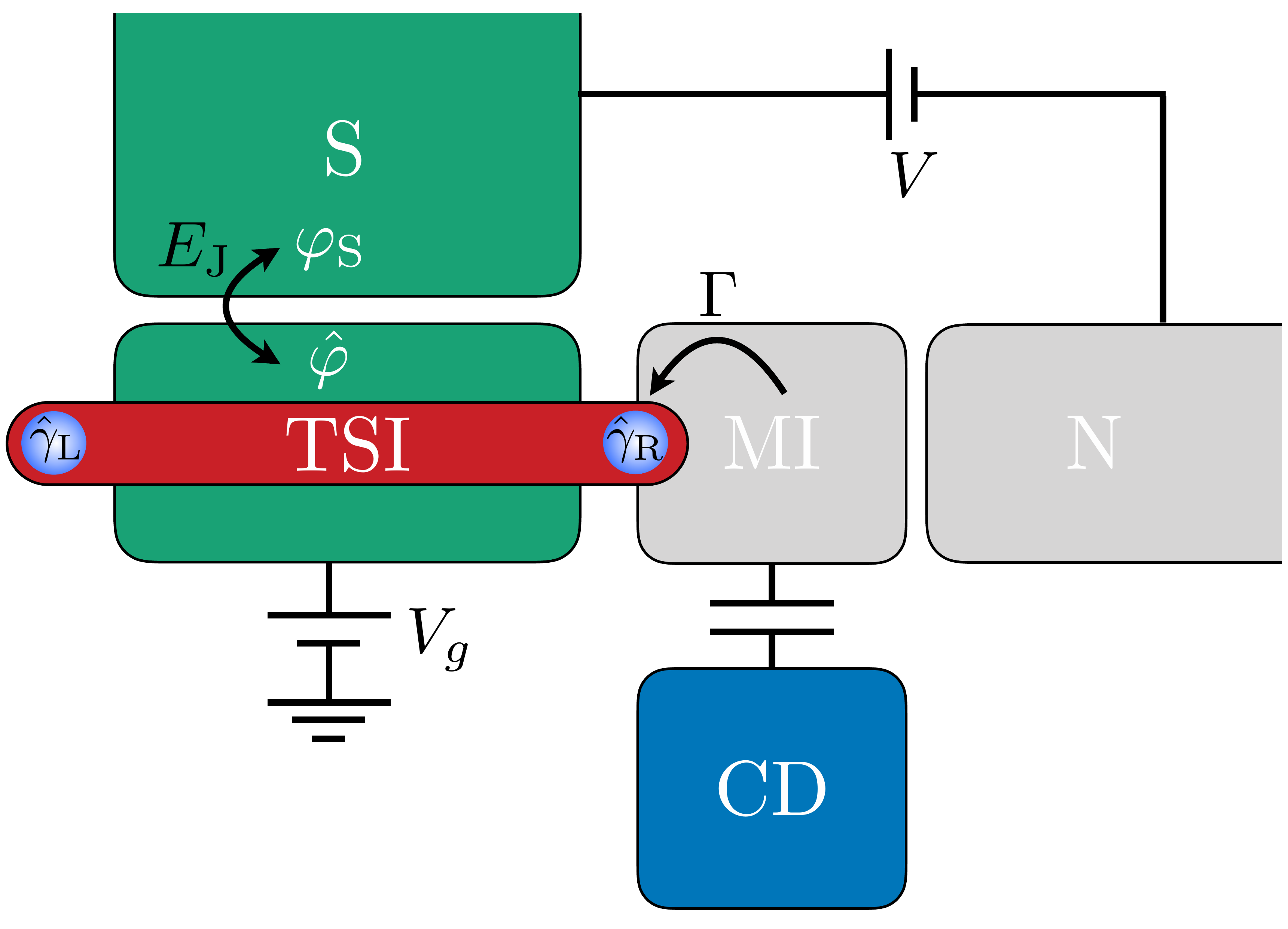}
	\caption{The Majorana single-charge transistor consists of a topological superconducting island (TSI) which can host Majorana bound states ($\hat{\gamma}_\text{L,R}$) and Cooper pairs. It is tunnel coupled ($\Gamma$) to a normal (N) and Josephson coupled ($E_\text{J}$) to a superconducting lead (S). A bias voltage $V$ is applied between them, and a gate voltage $V_g$ tunes the electrostatic potential of the island. To access the full counting statistics of single-charge tunneling, a metallic island (MI) is defined and read out using a charge detector (CD).
			 }
	\label{fig:1}
\end{figure}

In this work, we study electron transport through a \textit{Majorana single-charge transistor}.
The device consists of a topological superconducting island (TSI) --- which hosts Majorana quasiparticles as well as Cooper pairs ---  contacted by one normal (N) and one superconducting lead (S), see Fig.~\ref{fig:1}.
Both normal and anomalous single-charge tunneling between island and normal lead is facilitated by the Majorana bound state, while the Josephson coupling between island and superconductor allows for the transfer of two charges in form of a Cooper pair.
The Coulomb interaction on the island strongly influences the electric current~\cite{huetzen_2012,didier_2013,vanBeek_2016}.
At low bias voltage, transport is dominated by the so-called Josephson-Majorana cycle~\cite{didier_2013}, a sequence of normal and anomalous single-charge and Josephson tunneling. 
In the present paper, we go beyond the analysis of the average electron current.
We, rather, study the full counting statistics of all tunneling events as well as the distribution of waiting times between two charge-transfer events~\cite{fu_2022}. To enable a time-resolved measurement of the individual tunneling events, we propose to insert a small metallic island (MI) between the transistor and the normal lead, which can then be read out in real time by an electrostatically coupled charge detector (CD) such as a single-electron transistor~\cite{maisi_2011,maisi_2014}. 

This paper is organized as follows. In Sec.~\ref{sec:system}, we introduce the Hamiltonian of the Majorana single-charge transistor. Then, in Sec.~\ref{sec:transport}, we use real-time diagrammatic technique to study the non-equilibrium charge dynamics, in particular the Josephson-Majorana cycle, which involves the Majorana bound states and Cooper pairs.
In Sec.~\ref{sec:fcs}, we analyze the full counting statistics of electron tunneling.
We identify the highly-correlated nature of the charge transfers within the Josephson-Majorana cycle.
Furthermore, we predict a topological transition in the full counting statistics. Finally, in Sec.~\ref{sec:conclusions}, we conclude our findings.
%%%%%%%%%%%%%%%%%%%%%%%%%%%%%%%%%%%%%%%%%%%%%%%%%%%%%%%%%%%%%%%%%%%%%%%%%%%%%%%%%%%%%%%%%%%%%%%%%%%%
%%%%%%%%%%%%%%%%%%%%%%%%%%%%%%%%%%%%%%%%%%%%%%%%%%%%%%%%%%%%%%%%%%%%%%%%%%%%%%%%%%%%%%%%%%%%%%%%%%%%

\section{System}\label{sec:system} 
\begin{figure}[t]
	\includegraphics[width=.5\textwidth]{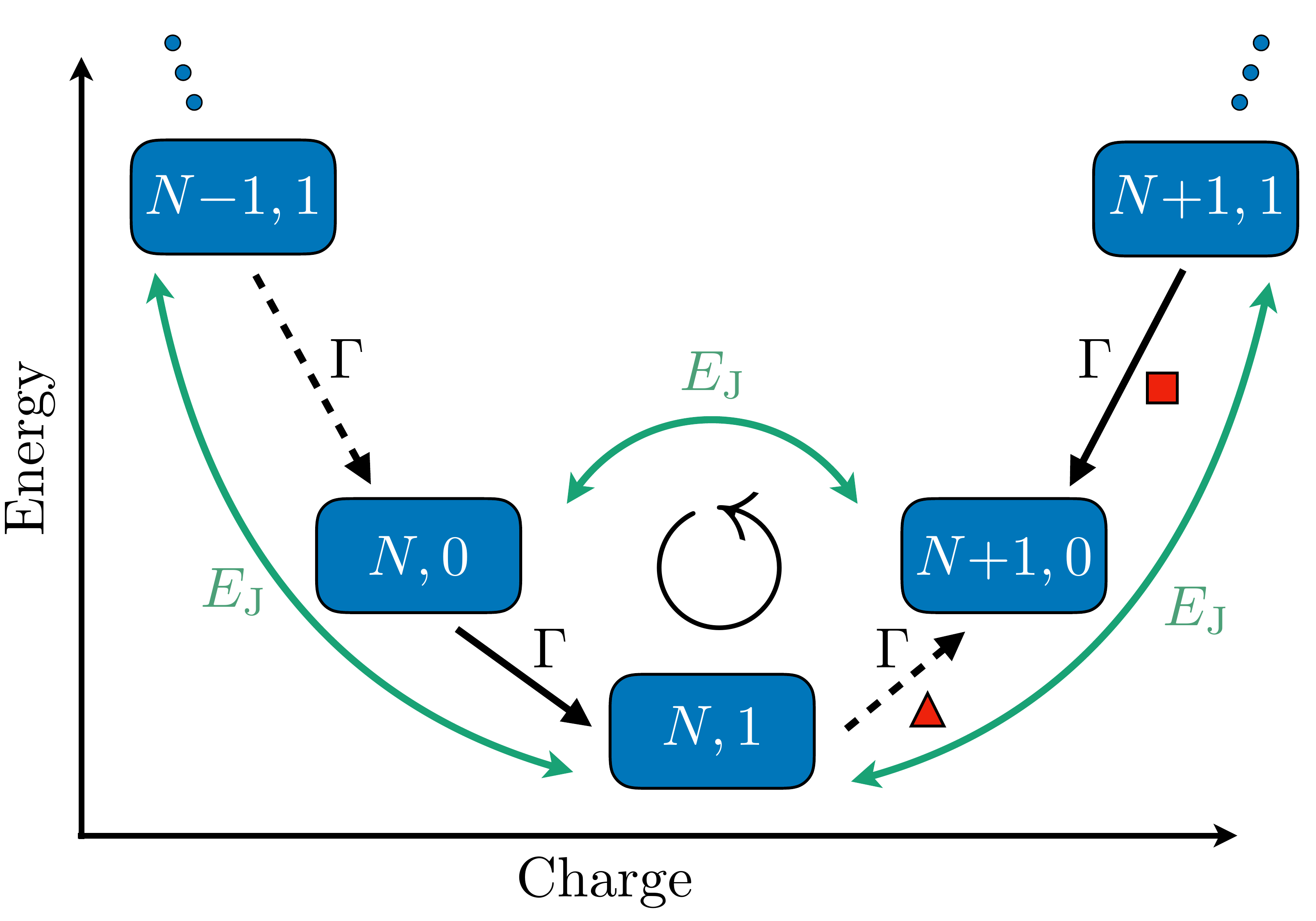}
	\caption{Relevant charge states $\ket{N,n}$ of the Majorana single-charge transistor. Solid lines indicate regular tunneling events and dashed lines anomalous ones that are present for bias voltages around $n_V=1$, for which the Josephson-Majorana cycle appears as shown. Green lines indicate the coherent transfer of Cooper pairs.
    For bias voltages around $n_V=0$, the direction of the arrow between $\ket{N,1}$ and $\ket{N+1,0}$ (marked by the triangle) changes, while for bias voltages around $n_V=2$, the direction of the arrow between $\ket{N+1,0}$ and $\ket{N+1,1}$ (marked by the square) is reversed.
    }\label{fig:2}
\end{figure}

The degrees of freedom of the TSI forming the central island of the transistor are the Majorana quasiparticles and the Cooper pairs. 
The Majorana bound states residing at the left and right end are described by operators $\hat\gamma_i =\hat \gamma_i^\dagger$ with $i=\text{L,R}$ that fulfill the anticommutation relations $\{\hat\gamma_i,\hat\gamma_j\} = \delta_{ij}$, i.e., the Majorana quasiparticles are their own antiparticles and they are neither bosons nor fermions. 
The Cooper pairs residing in the TSI are described by the number operator $\hat N$ or, equivalently, its canonically conjugated partner, the phase operator $\hat\varphi$. 
They fulfill the canonical commutation relation $[\hat N,e^{\pm i\hat \varphi}]= \pm e^{\pm i\hat \varphi}$, such that $e^{\pm i\hat \varphi}$ changes the number of Cooper pairs on the TSI by $\pm 1$.
We model the Majorana single-charge transistor by the Hamiltonian
\begin{equation}
    {H} = {H}_\text{C} + {H}_\text{J} + {H}_\text{N} + {H}_\text{T},
\end{equation}
which contains four parts.
The first one,
\begin{equation}
    {H}_\text{C} = E_\text{C} (2\hat{N} + \hat{d}^\dagger \hat{d} - n_g)^2,
\end{equation}
accounts for the Coulomb interaction of all charges on the TSI, which is assumed to be large due to the  mesoscopic scale of the island.
Here, $E_\text{C}$ defines the energy scale of the charging energy.
The total charge that enters this expression is given by twice the number of Cooper pairs, a non-local electron (if present) residing in the zero-energy level comprised from the two Majorana bound states, and the gate charge $n_g$ that can be tuned by the gate voltage $V_g$.
The annihilation operator for the non-local electron state can be expressed as
\begin{align}
\hat d=\frac{1}{\sqrt{2}}e^{-i\frac{\hat \varphi}{2}} \left(\hat\gamma_\text{L} + i \hat\gamma_\text{R}\right) \, .
\end{align}
The associated number operator $\hat n=\hat{d}^\dagger \hat{d}$ can conveniently be used to define the charge eigenstates $\ket{N,n}$ of the island with $N$ being the number of Cooper pairs and $n=0,1$ denoting the occupation of the zero-energy level~\cite{fu_2010,van_heck_2011,zazunov_2012}.
For the rest of the paper, we choose $n_g=2N+1$ for some arbitrarily chosen number $N$ of Cooper pairs, such that the ground state is $\ket{N,1}$ and the first two excited states, $\ket{N,0}$ and $\ket{N+1,0}$, are energetically degenerate, see Fig.~\ref{fig:2}.
This is the situation, in which the Josephson-Majorana cycle occurs~\cite{didier_2013}.

The second term of the Hamiltonian,
\begin{equation}
    {H}_\text{J} = -E_\text{J} \cos (\hat{\varphi}-\varphi_\text{S} ),
\end{equation}
describes the Josephson coupling of the TSI to a bulk superconducting lead at zero electrochemical potential, $\mu_\text{S}=0$.
It allows for a coherent exchange of Cooper pairs, $\ket{N,n}\leftrightarrow \ket{N\pm1,n}$, see Fig.~\ref{fig:2}. Here, $E_\text{J}$ is the coupling energy and $\varphi_\text{S}$ is the phase of the bulk superconductor. 
While the phase of the superconducting lead is fixed, both the phase ($\hat \varphi$) and Cooper pair number ($\hat N$) of the TSI fluctuate. Moreover, we assume the superconducting gap $\Delta$ to be sufficiently large such that the involvement of Bogoliubov quasiparticles --- so called quasiparticle poisoning~\cite{shaw_2008,plugge_2016,karzig_2021} --- can be neglected. 
A coupling from the superconductor to the Majorana mode can be safely ignored~\cite{zazunov_2012}. 

The third term of the Hamiltonian models a normal conducting lead of noninteracting electrons, ${H}_\text{N} =  \sum_k \epsilon_k  \hat{c}^\dagger_{k} \hat{c}_{k}$, with a Fermi-Dirac occupation $\ev{\hat{c}^\dagger_{k} \hat{c}_{k}}=f(\epsilon_k-\mu_\text{N})$, determined by the energy $\epsilon_k$ and the electrochemical potential $\mu_\text{N}=eV$. 
Here, the bias voltage $V$ between normal and superconducting lead enters.

Finally, tunneling between normal lead and TSI is given by the fourth term of the Hamiltonian.
It can be written in the form~\cite{zazunov_2011,didier_2013}
\begin{align}
{H}_\text{T}= \sum_k t_{k}\left(\hat{d}^\dagger - e^{i\hat{\varphi}} \hat{d} \right)\hat{c}_k  + \text{h.c.},
\end{align}
which couples electrons in the lead $\hat c_k$ to the right Majorana quasiparticle $e^{i\hat{\varphi}/2}\hat{\gamma}_\text{R}\sim \left(\hat{d}^\dagger - e^{i\hat{\varphi}} \hat{d}\right)$ with tunneling amplitude $t_k$, which we assume to be the same for all lead states, $t_k=t$.
It is important to note that there are two qualitatively different terms contributing to tunneling.
In addition to normal tunneling ($\sim \hat  d^\dagger \hat c_k$) which simply transfers an electron from the lead to the non-local electron state (or vice versa), there is also anomalous tunneling ($\sim e^{i\hat \varphi}\hat d \,\hat c_k$) where a Cooper pair is formed via the annihilation of an electron in the lead and the non-local electron of the island (or vice versa).

We remark that if Bogoliubov quasiparticles were involved, we would have to take additional tunneling events coupling even- and odd-parity states into account. 
Furthermore, we emphasize that in the system we study, only one of two Majorana quasiparticles is involved. 
As a consequence, the non-local character of the Majorana bound state is not directly probed in the proposed transport geometry.
Therefore, our results are not suited to distinguish a Majorana bound state from an accidental zero-energy Andreev bound state~\cite{yavilberg_2019,prada_2020}.

In order to monitor in time the single-charge tunneling events, the normal lead is supposed to be separated into a metallic island MI with finite charging energy and a macroscopic normal conductor N.
Then, the charge state of the MI can be measured by a capacitively-coupled charge detector CD, see Fig.~\ref{fig:1}.
For a sufficiently large bias voltage the current through the MI is unidirectional, and one can uniquely distinguish tunneling between TSI and MI from tunneling between MI and N.

To take the separation of the normal lead into MI and N into account, we simply interpret $H_\text{N}$ as the Hamiltonian for the MI and neglect any voltage drop between MI and N (which is justified if the coupling between TSI and MI is much weaker than between MI and N).

\section{Electronic transport}\label{sec:transport}
To drive a current through the island, a bias voltage $\mu_\text{N}-\mu_\text{S}=eV$ is applied between the normal and superconducting lead. For convenience, we introduce the dimensionless bias voltage $n_V=eV/(2E_\text{C})$. In order to describe the non-equilibrium dynamics, we employ the real-time diagrammatic technique~\cite{koenig_1996_1,koenig_1996_2} which is a systematic perturbation expansion in the tunnel coupling strength $\Gamma={2\pi}\vert t\vert^2D(\epsilon_F)/\hbar$, where $D(\epsilon_F)$ is the density of states  at the Fermi energy $\epsilon_F$ in the metallic reservoir. Formally, we obtain in leading-order perturbation theory (justified for $\Gamma \ll k_\text{B} T$) with the Markov approximation the generalized master equation for the reduced density matrix~\cite{kleinherbers_2020}
\begin{align}\label{eq:mastereq}
\dot{\rho}={\cal L}\rho=-\frac{i}{\hbar} \left[ H_\text{C}+H_\text{J},\rho \right]+ {\cal W}\rho,
\end{align}
where ${\cal L}$ is the full Liouvillian. The superoperator ${\cal W}$ describes generalized tunneling rates of the form (for details see Appendix~\ref{details})
\begin{align}\label{eq:rate}
\Gamma_\pm(\Delta E)=\frac{\Gamma}{2}\left[f_\pm(\Delta E-\mu_\text{N}) + i R(\Delta E-\mu_\text{N})\right],
\end{align}
where $\Delta E$ are the excitation energies of the system.
Here, the real part given by the Fermi-Dirac function $f_+(x)=f(x)$ and $f_-(x)=1-f(x)$ describes the relaxation dynamics and the imaginary part $\pi R(x)=\Re\left[\psi\left(\frac{1}{2} + i\frac{x}{2\pi k_\text{B}T} \right)\right]-\ln\left( \frac{W_c}{{2\pi k_\text{B} T}}\right)$ given by the digamma function $\psi(x)$~\cite{abramowitz_1988} modifies the coherent dynamics of the system. The cutoff parameter $W_c\gg k_\text{B} T$ is used for regularization but drops out exactly for all quantities shown in the figures. Note that Eq.~\eqref{eq:mastereq} is equivalent to the Redfield equation~\cite{timm_2008}. 

To solve the master equation, we have to truncate the infinite Hilbert space to a relevant and finite subspace. For this purpose, we choose as a basis the three odd-parity states $\ket{N-1,1},\ket{N,1},\ket{N+1,1}$ as well as the two even-parity states $\ket{N,0},\ket{N+1,0}$, see Fig.~\ref{fig:2}. Hence, we cut off all higher-energy states, which is justified if the normal metal can neither thermally ($k_\text{B} T$) nor via the bias voltage ($n_V$) excite them. This leads us to the following hierarchy of energy scales
\begin{align}
\Gamma\ll k_\text{B} T\ll E_\text{C}\ll \Delta
\end{align}
as well as the restriction for the bias voltage
\begin{align}
|n_V|\lesssim \frac{5}{2}.
\end{align}
Furthermore, to exclude a coherent transition to higher-energy states via the Josephson coupling, we additionally assume
\begin{align}
E_\text{J}\ll E_\text{C} \, ,
\end{align}
but we do not specify the relative size of $E_\text{J}$ to $\Gamma$.
In fact, we will later see that the systems behaves qualitatively differently in the regimes $E_\text{J} \ll \Gamma$ and $E_\text{J} \gg \Gamma$, respectively.
For the above specified hierarchy of energy scales, where we can truncate the Hilbert space to a five-dimensional subspace (cf. Fig.~\ref{fig:2}), the eigenstates of $H_\text{C}+H_\text{J}$ are given by
\begin{align}
\ket{\Psi_{1}}&=\cos \theta \ket{N,1} + \frac{\sin \theta}{\sqrt{2}}\left(\ket{N-1,1}+\ket{N+1,1}\right), \nonumber\\
\ket{\Psi_{2}}&=-\sin \theta \ket{N,1} + \frac{\cos \theta}{\sqrt{2}}\left(\ket{N-1,1}+\ket{N+1,1}\right), \nonumber\\
\ket{\Psi_{3}}&=\frac{1}{\sqrt{2}}\left(\ket{N,0}+\ket{N+1,0}\right), \nonumber \\
\ket{\Psi_{4}}&=\frac{1}{\sqrt{2}}\left(\ket{N,0}-\ket{N+1,0}\right), \nonumber\\
\ket{\Psi_{5}}&= \frac{1}{\sqrt{2}}\left(\ket{N-1,1}-\ket{N+1,1}\right) \label{eq:eigenstates},
\end{align}
with the mixing angle defined by $\cos^2 \theta=\frac{1}{2}{+}\frac{\sqrt{2}E_\text{C}}{\sqrt{8E_\text{C}^2+E_\text{J}^2}}$. The corresponding eigenenergies are
\begin{align}
&E_{1,2}=2E_\text{C} \mp \frac{\sqrt{8E_\text{C}^2+E_\text{J}^2}}{\sqrt{2}}, \nonumber\\
&E_{3,4}=E_\text{C}\mp\frac{E_\text{J}}{2}, \nonumber\\
&E_5=4E_\text{C},
\end{align}
where $\ket{\Psi_1}$ is the ground state.

\begin{figure}[t]
	\includegraphics[width=.5\textwidth]{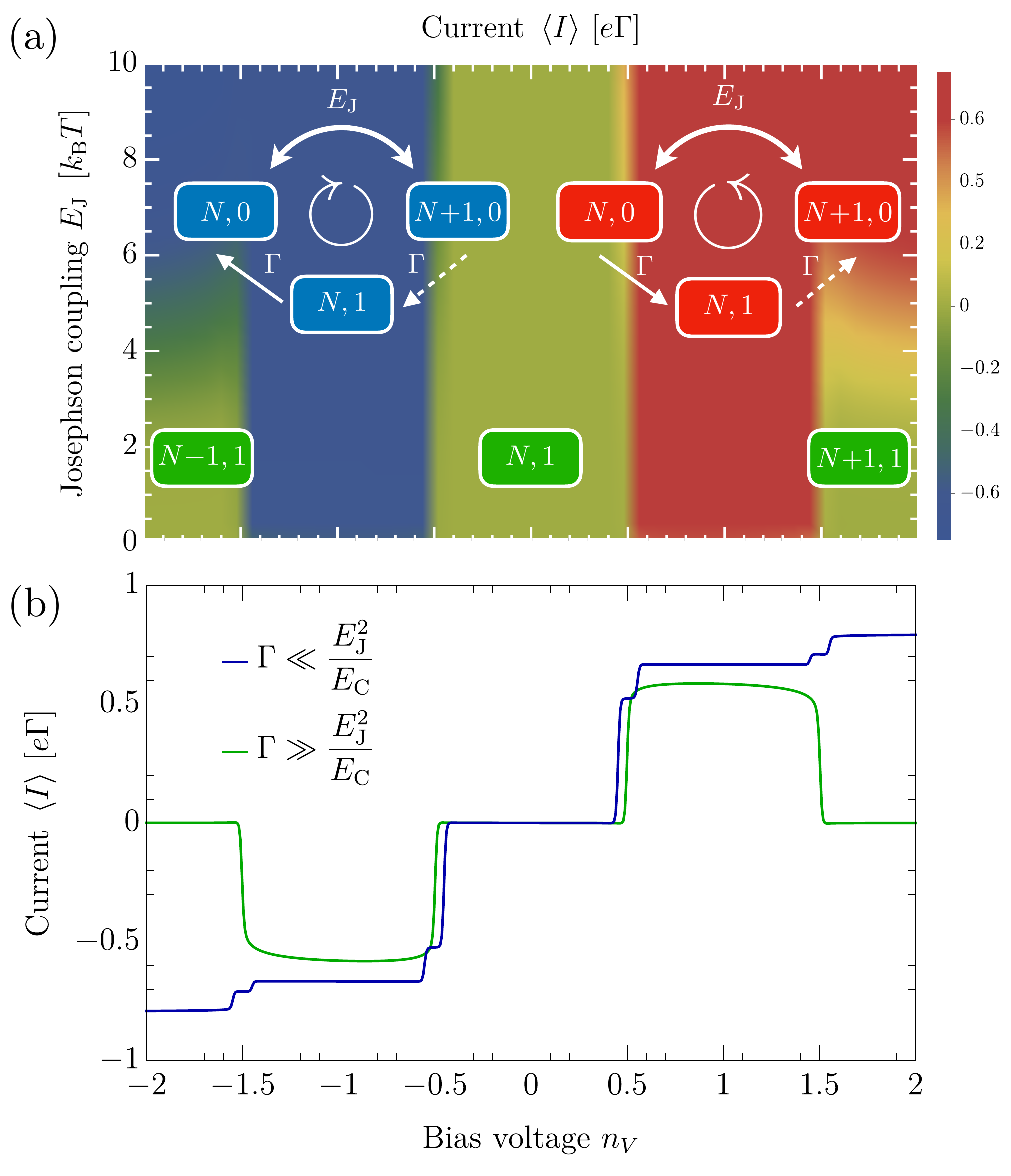}
	\caption{(a) Current $\ev{I}$ through the single-charge transistor as a function of the Josephson coupling $E_\text{J}$ and the bias voltage $n_V$. For $1/2 \lesssim  \vert n_V\vert  \lesssim 3/2$, the relevant transport mechanism, the Josephson-Majorana cycle, is indicated. In the green regions, the transport is blocked and the system is stuck in a single state. In (b), we choose the specific values $E_\text{J}=0.2 \,k_\text{B} T$ (green) and  $E_\text{J}=20 \,k_\text{B} T$ (blue) for the Josephson coupling. The remaining parameters are
    $\Gamma=0.1\, k_\text{B} T$ and $E_\text{C}=100 \,k_\text{B} T$.
	}
\label{fig:3}	
\end{figure}

The net electron current can be determined via 
\begin{align}\label{eq:current}
\ev{I}= e \tr \left[\left( {\cal J}_+ -{\cal J}_-\right)  \rho_\text{st}\right],
\end{align}
where the jump operators ${\cal J}_+$ and ${\cal J}_-$ describe all tunneling events where an electron enters and leaves the TSI via the normal metal, respectively (for details see Appendix~\ref{details}). The stationary state is defined by ${\cal L}\rho_\text{st}=0$. In Fig.~\ref{fig:3}(a), the current $\ev{I}$ through the device is shown as a function of the bias voltage $n_V$ and the Josephson coupling $E_\text{J}$. 
We can identify different transport regimes. 

(i) For $ \vert n_V \vert  \lesssim 1/2$, we are in the Coulomb blockade regime (green) where only the ground state $\ket{N,1}$ is occupied.
As a consequence, transport is blocked. 

(ii) For intermediate bias voltages $1/2 \lesssim  \vert n_V\vert  \lesssim 3/2$, there are now three states accessible, namely $\ket{N,1}$, $\ket{N,0}$ and $\ket{N+1,0}$. In this regime, the transport is possible via the Josephson-Majorana cycle~\cite{didier_2013} which for $n_V>0$ (red) corresponds to a cyclic repetition of a normal tunneling event, $\ket{N,0}\rightarrow \ket{N,1}$, an anomalous tunneling event, $\ket{N,1}\rightarrow \ket{N+1,0}$, and the coherent transfer of a Cooper pair $\ket{N+1,0}%\leftrightarrow
\rightarrow\ket{N,0}$. For $n_V<0$ (blue) the direction of the cycle is reversed, see the insets of Fig.~\ref{fig:3}(a). 

(iii) Finally, for larger bias voltage $3/2 \lesssim  \vert n_V\vert \lesssim 5/2$ all five states considered in this paper are accessible. Typically, the increase of number of accessible states opens new transport channels and, therefore, enhances the current. This happens also here in the case of a large Josephson coupling. For small Josephson coupling, however, the opposite effect occurs: transport becomes blocked. The reason is that the system becomes coherently trapped in the dark state $\ket{N+1,1}$ for $n_V>0$ and $\ket{N-1,1}$ for $n_V<0$. The only possible escape from this dark state is to enter the ground state $\ket{N,1}$ via an exchange of a Cooper pair. However, this coherent process is suppressed due to an energy difference $\sim 4 E_\text{C}$ of the states. 

For regimes (ii) and (iii), we are able to provide approximate analytical expressions for the current by using the approximate eigenstates of Eq.~\eqref{eq:eigenstates} with $\theta=0$. 
For $n_V=1$, regime (ii), we obtain
\begin{align}
\ev{I}=\frac{2e \Gamma}{3+\left(\frac{\Gamma}{E_\text{J}}\right)^2},
\end{align}
where we neglected renormalization $R(x)\rightarrow 0$ and replaced the Fermi function by the Heaviside function, $f(x)=\Theta(-x)$.
We observe that the current via the Josephson-Majorana cycle is independent of the charging energy $E_\text{C}$ and, therefore, not affected by Coulomb-blockade.

For $n_V=2$, regime (iii), we similarly find 
\begin{align}
\ev{I}=\frac{4e \Gamma}{5+64\left(\frac{\Gamma E_\text{C}}{E_\text{J}^2}\right)^2+2\left(\frac{\Gamma}{E_\text{J}}\right)^2}.
\end{align}
Now, the current does depend on the charging energy $E_\text{C}$. In fact, we can identify a criterion for a blocked current 
\begin{align}
\Gamma \ll  \frac{{E_\text{J}^2}}{E_\text{C}}&: \quad \ev{I}\approx \frac{4}{5} e\Gamma, \\
\Gamma \gg  \frac{{E_\text{J}^2}}{E_\text{C}}&: \quad \ev{I}\approx \frac{e \Gamma}{16} \left( \frac{E_\text{J}^2}{\Gamma E_\text{C}}\right)^2 \, .
\end{align}
In the former case, the current is increased as compared to $n_V=1$, while in the latter case it is algebraically suppressed by Coulomb repulsion $E_\text{C}$.
For illustration, in Fig.~\ref{fig:3}(b), we show the current in both cases as a function of the bias voltage $n_V$. 
For $\Gamma \ll  \frac{{E_\text{J}^2}}{E_\text{C}}$ (blue), we get the expected result, where the current increases stepwise with $n_V$ as soon as the number of excitation energies in the bias window increases.
In contrast, for $\Gamma \gg  \frac{{E_\text{J}^2}}{E_\text{C}}$ (green), we observe a negative differential conductance because the current decreases although the bias voltage is increased. The algebraic suppression of the current is a signature of the dark state due to coherent Coulomb blockade. 

\section{Real-time analysis of charge transfer}\label{sec:fcs}
In the following, we study the real-time statistics of single-charge tunneling from the normal metal into the TSI in the parameter regime of the Josephson-Majorana cycle, i.e., around $n_V=1$. 
To access this statistics experimentally, we propose a setup as shown in Fig.~\ref{fig:1}, where the normal metal is divided by a tunneling barrier into a small metallic island (MI) and a bulk metallic lead (N). Then, the total charge of the metallic island can be read out using a nearby charge detector (CD), e.g., a single-electron transistor~\cite{maisi_2011,maisi_2014}. 
The resulting statistics of tunneling events will be analyzed using waiting times and full counting statistics. 

For $n_V=1$, it is sufficient to consider only the three states $\ket{N,1},\ket{N,0}$, and $\ket{N{+}1,0}$.
Then, the Liouvillian, which acts on the density matrix $\rho=(\rho_{N,1},\rho_{N,0},\rho_{N{+}1,0},\rho_{N{+}1,0}^{N,0},\rho_{N,0}^{N{+}1,0})$ is described by a $5\times 5$ matrix.
The first three elements of $\rho$ are populations $\rho_\chi=\mel{\chi}{\rho}{\chi}$ and the last two elements are coherences $\rho_{\chi^\prime}^{\chi}=\mel{\chi}{\rho}{\chi^\prime}$.
Including counting variables $z_n$ and $z_a$ to count normal and anomalous tunneling events into the TSI, we find
\begin{align}\label{eq:liouvillian_cycle}
{\cal L}_{z_n,z_a}=\begin{pmatrix}
 -\Gamma  &  z_n\Gamma & 0   & 0 & 0 \\
0  & -\Gamma & 0 & -\frac{i E_\text{J}}{2} & \frac{i E_\text{J}}{2} \\
z_a \Gamma  & 0 & 0  & \frac{i E_\text{J}}{2} & -\frac{i E_\text{J}}{2} \\
 0 & -\frac{i E_\text{J}}{2} & \frac{i E_\text{J}}{2} & -\frac{\Gamma }{2} -i \omega & 0 \\
 0 & \frac{iE_\text{J}}{2} & -\frac{i E_\text{J}}{2} & 0 & -\frac{\Gamma }{2}+i \omega \\
\end{pmatrix},
\end{align}
where we replaced the Fermi function by the Heaviside function, $f(x)=\Theta(-x)$, and included renormalization effects in the energy $\omega$ which is obtained via
\begin{align}
\omega\approx &\frac{\Gamma}{2} \sum_\pm \pm R(\mp E_\text{C}-\mu_\text{N})\pm R(\mp 3E_\text{C}-\mu_\text{N})\approx \Gamma \frac{\ln 15}{2\pi}.
\end{align}
It originates in those diagrams describing virtual charge fluctuations from $\ket{N,0}$ to $\ket{N{-}1,1}$ and $\ket{N,1}$ as well as from  $\ket{N{+}1,0}$ to $\ket{N,1}$ and $\ket{N{+}1,1}$.

The Liouvillian entering Eq.~\eqref{eq:mastereq} is obtained by setting $z_n=z_a=1$, i.e. ${\cal L}={\cal L}_{1,1}$.
The jump operators for normal and anomalous tunneling into the TSI are, then, given by
\begin{align}
{\cal J}_n=\partial_{z_n}{\cal L}_{z_n,1},\\
{\cal J}_a=\partial_{z_a}{\cal L}_{1,z_a}, 
\end{align}
which add up to the total jump operator ${\cal J}={\cal J}_n+{\cal J}_a$ for tunneling into the TSI.

\begin{figure}[t]
	\includegraphics[width=.5\textwidth]{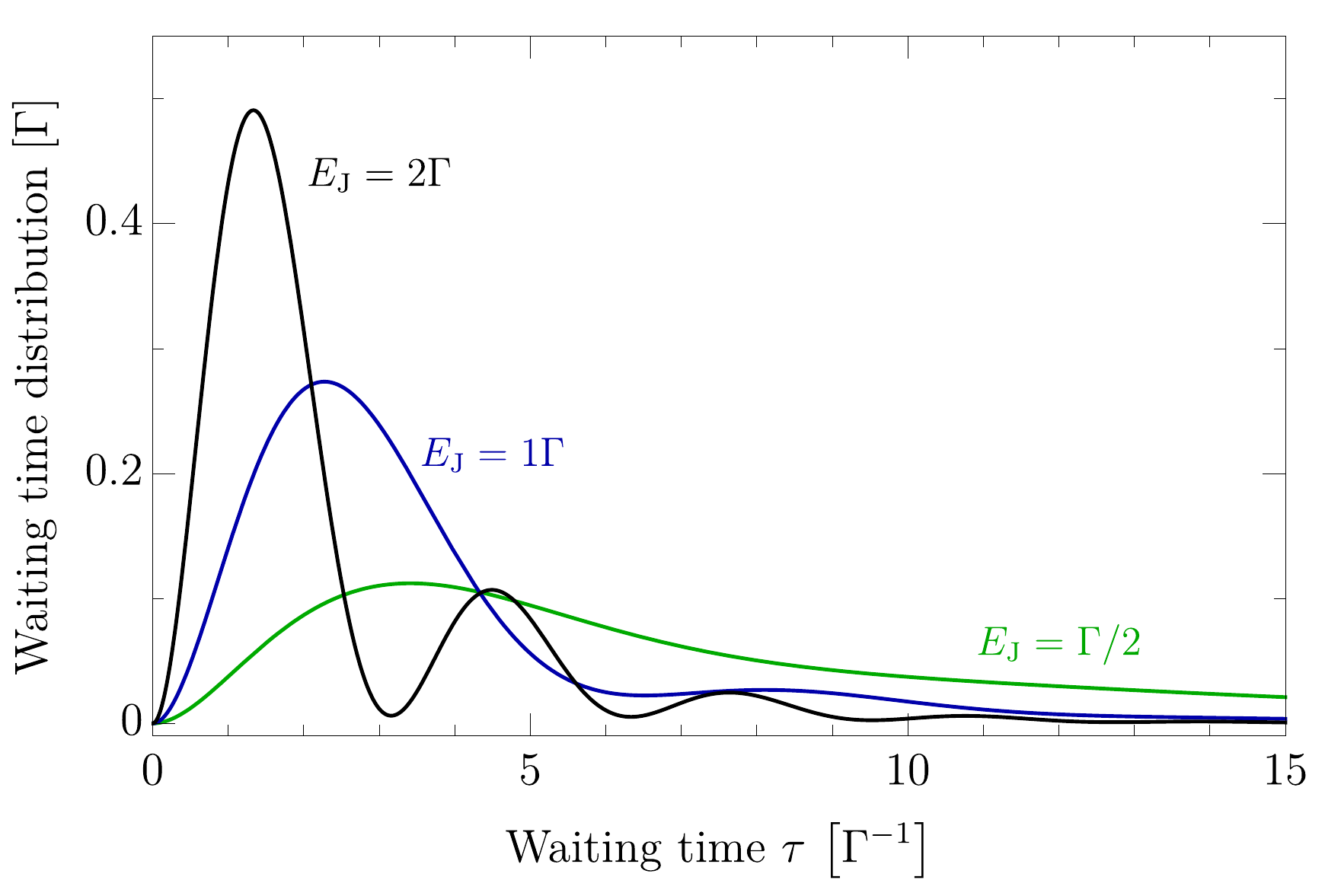}
	\caption{Waiting time distribution between successive anomalous and normal tunneling events with $E_\text{J}=2\Gamma$ (black), $E_\text{J}=1\Gamma$ (blue), and $E_\text{J}=\Gamma/2$ (green). 
 }
	\label{fig:waiting_times}
\end{figure}

\subsection{Waiting times}
One way to characterize the tunneling statistics is the waiting-time distribution $w(\tau)$, defined as the distribution of the times $\tau$ between successive anomalous and normal tunneling events, i.e., the time the system dwells in the even-parity states $\ket{N,0}$, and $\ket{N{+}1,0}$.
For this, we calculate 
\begin{align}
w(\tau)&=\frac{\tr\left[ {\cal J}_n e^{\left( {\cal L}-{\cal J}_n-{\cal J}_a\right)\tau} {\cal J}_a \rho_\text{st}\right]}{\tr\left[ {\cal J}_a \rho_\text{st}\right]} \, .
\label{eq:wt}
\end{align} 
In Fig.~\ref{fig:waiting_times}(a), we show the waiting time distribution $w(\tau)$ for three different values of the Josephson coupling $E_\text{J}$. 
For $E_\text{J}=2\Gamma$ (black), the waiting time distribution shows clear oscillations which indicate the coherent transfer of Cooper pairs back and forth between the system and the superconducting lead. For $E_\text{J}=\Gamma$ (blue), the oscillations are less visible. If we go to even smaller Josephson couplings $E_\text{J}=\Gamma/2$ (green), the oscillations completely disappear and the waiting times become longer.
In fact, a bottleneck is created, where the waiting times depend only on the time required for the transfer of a Cooper pair into the superconductor. Coherent oscillations do not occur because a new electron tunnels in from the normal metal almost immediately.

\subsection{Full counting statistics}
For a more complete picture, we study the full counting statistics of tunneling events. 
In the following, we no longer differentiate between normal and anomalous tunneling events but count both of them on equal footing.
This is achieved by setting $z=z_n=z_a$, i.e., by using the Liouvillian ${\cal L}\z={\cal L}_{z,z}$. Then, the probability that $M$ tunneling events (either normal or anomalous tunneling) happen in a time interval of length $t$ is given by
\begin{align}
P_M(t)=\frac{1}{M!} \partial_z^M\tr\left( e^{{\cal L}\z t}\rho_\text{st}\right)\vert_{z=0}.
\end{align}
For each interval length $t$, there is a distribution $P_M(t)$ of the number $M$ of counted tunneling events.
While for continuous stochastic variables, probability distributions are naturally characterized by ordinary cumulants, in the case of discrete stochastic variable, such as the number of tunneling events, it is more natural to use \textit{factorial cumulants}. 
They are conveniently derived via 
\begin{align}
C_{\text{F},m}=\partial_z^m {\cal S}(z,t)\vert_{z=1},
\end{align}
with the cumulant-generating function 
\begin{align}
{\cal S}(z,t)= \ln \tr\left( e^{{\cal L}\z t}\rho_\text{st}\right).
\end{align}
The first factorial cumulant is the mean value of the number of single-charge tunneling events, $C_{\text{F},1}=\ev{M}$, and the second one is a particular linear combination of the variance and the mean value, $C_{\text{F},2}=\ev{M^2}-\ev{M}^2-\ev{M}$. 
It is obvious that the variance, and thus $C_{\text{F},2}$, contains extra information of the distribution in addition to the mean value.
In analogy, higher-order factorial cumulants $C_{\text{F},m}
$ with $m>2$ reveal more and more information about the distribution.

Using factorial instead of ordinary cumulants to characterize the full counting statistics of the single-charge tunneling has several advantages.
First, they are more natural for discrete stochastic variables~\cite{koenig_2021newton}.
Second, factorial cumulants avoid the unwanted feature of universal oscillations~\cite{flindt2009universal} and, thus, are useful tools to identify system-specific informations. 
Third, higher-order ($m>1$) factorial cumulants show an intrinsic resilience to detection errors such as a finite time resolution or false noise-induced events~\cite{kleinherbers_2021_pushing}. 
Finally, they are useful to identify correlations between the tunneling events~\cite{schomerus_2001,kambly_2011,stegmann2015detection,kleinherbers2018revealing,kurzmann_2019}.
If all the tunneling events were uncorrelated and occurred with the same single-particle tunneling probability, then the full counting statistics would be described by a Poisson distribution, and all higher-order factorial cumulants would vanish.
The more general scenario of uncorrelated tunneling events with non-identical tunneling probabilities yields a Poisson-binomial distribution~\cite{wang_1993}, for which all factorial cumulants are, in general, non-zero.
However, as explicitly shown in  Refs.~\cite{kambly_2011,stegmann2015detection}, the sign of the factorial cumulants is, in the case of uncorrelated tunneling, fixed and given by
\begin{align}\label{eq:interactions_sign}
(-1)^{m-1}C_{\text{F},m}\ge 0.
\end{align}
This allows us to define quite a strong criterion for the presence of correlations: whenever Eq.~\eqref{eq:interactions_sign} is violated for any order $m$, any interval length $t$ and any set of system parameters, then the distribution $P_M$ cannot be written as a Poisson binomial distribution, i.e., the tunneling events are correlated.

For $m=2$, the criterion for the presence of correlation is equivalent to a super-Poissonian Fano factor
\begin{align}\label{eq:signcrit}
-C_\text{F,2}< 0 \quad  \Leftrightarrow \quad F=\frac{\ev{M^2}-\ev{M}^2}{\ev{M}}> 1,
\end{align} 
which is a well-established quantity in the study of correlated electron transfer~\cite{cottet_2004,belzig_2005,koch_2005,zarchin_2007,kiesslich_2007}. 
But since Eq.~\eqref{eq:interactions_sign} should (for uncorrelated tunneling) hold for any order $m$, the  factorial cumulants generalize the criterion $F>1$ to a whole family of criteria to detect correlations.

\begin{figure}[t]
	\includegraphics[width=.5\textwidth]{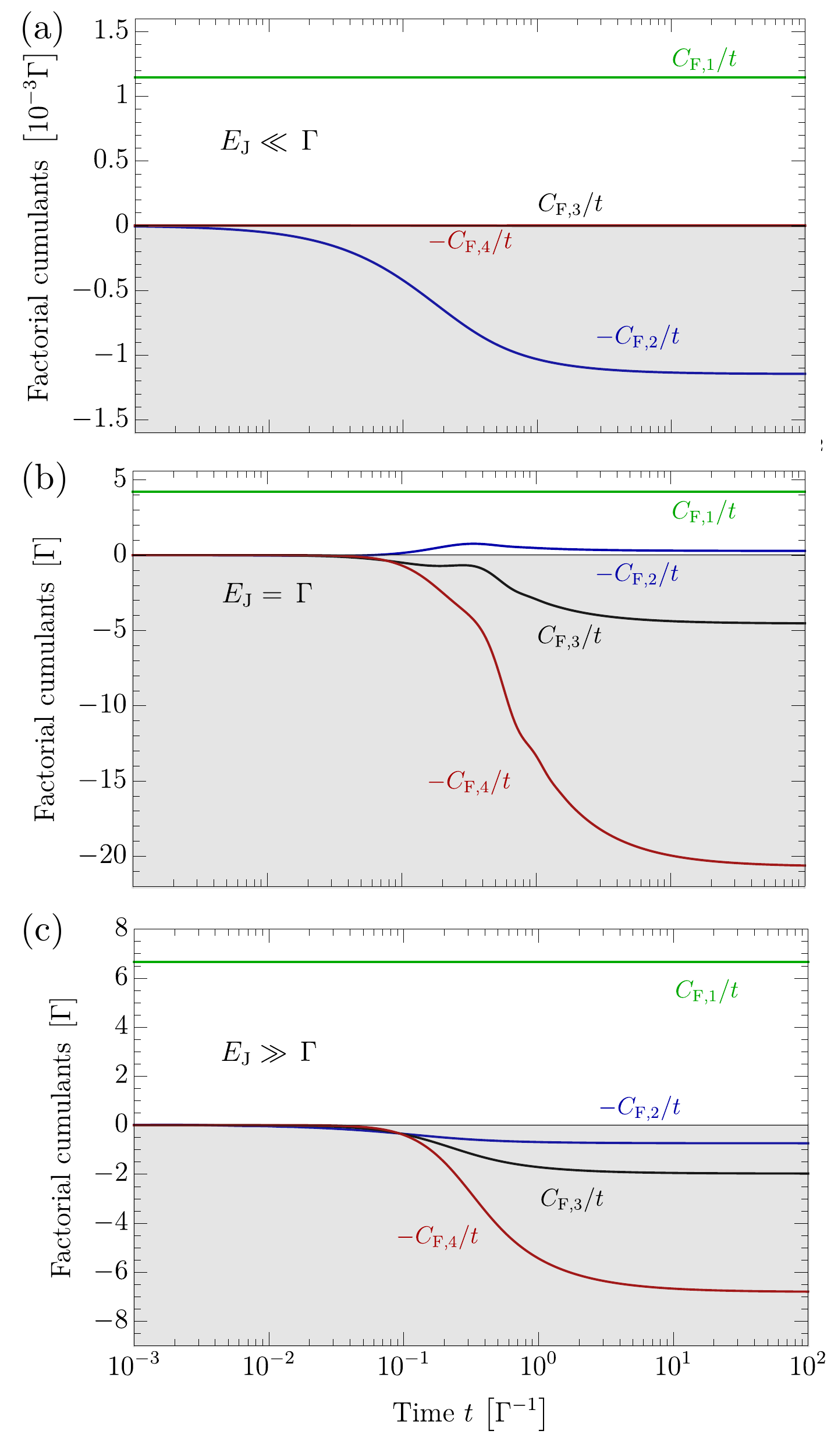}
	\caption{(a)-(c) First four factorial cumulants as a function of time for (a) $E_\text{J}=0.01\Gamma$, (b) $E_\text{J}=\Gamma$, and (c) $E_\text{J}=100\Gamma$. Cumulants in the gray shaded area indicate a negative sign of $(-1)^{m-1}C_{\text{F},m}$ and thus correlations. The charging energy is given by $E_\text{C}=1000\,\Gamma$.
 }
	\label{fig:facs}
\end{figure}

In Fig.~\ref{fig:facs}(a)-(c), we show the first four factorial cumulants $C_{\text{F},m}$ as a function of time $t$ for three different values of the Josephson coupling $E_\text{J}$. 
A violation of Eq.~\eqref{eq:signcrit}, i.e., the presence of correlations is indicated by factorial cumulants $(-1)^{m-1}C_{\text{F},m}$ entering the gray shaded area.
This is the case for all parameters, which means that the electron transfer within the Josephson-Majorana cycle is highly correlated.

In  Fig.~\ref{fig:facs}(a), we choose the Cooper-pair transfer to be the bottleneck, $E_\text{J}\ll \Gamma$.
We find that the distribution is well described by the first two factorial cumulants $C_\text{F,1}$ and $C_\text{F,2}$, while those of higher order are negligible. The Fano factor $F=1+(C_{\text{F},2}/C_{\text{F},1})$ changes as a function of the interval length $t$ from $F=1$ at short times, $\Gamma t\ll1$, to $F=2$ in the long-time limit, $\Gamma t\gg1$. 
Interpreting the Fano factor as an effective charge transferred in a Poisson-like processes, a Fano factor of $F=2$ indicates that in the long-time limit the full counting statistics effectively looks like a Poisson process of Cooper-pair transfers.
The fact that it is actually single-electron tunneling events that are counted is expressed in the Fano factor $F=1$ for short times.

In  Fig.~\ref{fig:facs}(b), for $E_\text{J}=\Gamma$, we observe a more complex structure of the factorial cumulants, albeit they still violate Eq.~\eqref{eq:signcrit} for the third and fourth factorial cumulant $C_{\text{F},3}$ and $C_{\text{F},4}$. 
Finally, in  Fig.~\ref{fig:facs}(c), for $E_\text{J}\gg \Gamma$, the bottleneck is due to the (normal and anomalous) single-charge tunneling events.
In this case, we can describe the dynamics by a simple rate equation (see Appendix~\ref{app:rate} for details)
\begin{align}\label{eq:switching}
\partial_t \begin{pmatrix}
 {p}_\text{odd}\\
{p}_\text{even}
\end{pmatrix}
=
\begin{pmatrix}
-\Gamma & z\frac{\Gamma}{2}\\
z \Gamma  &-\frac{\Gamma}{2}
\end{pmatrix}
\begin{pmatrix}
 {p}_\text{odd}\\
{p}_\text{even}
\end{pmatrix},
\end{align}
where $p_\text{odd}=\rho_{N,1}$ and $p_\text{even}=\rho_{N,0}+\rho_{N{+}1,0}$. Thus, the parity switches from even to odd via normal tunneling and from odd to even via anomalous tunneling. Each switch increases the electron counter. Thus, although the system is effectively described by a two-state model, the electron transport is correlated.
[If only normal or only anomalous tunneling were counted, the resulting full counting statistics would always fulfill Eq.~\eqref{eq:interactions_sign}.]

While the factorial cumulants provide via Eq.~\eqref{eq:interactions_sign} an experimentally accessible tool to detect the presence of correlations, there is, from a purely theoretical point of view, a more direct way to prove that the single-charge transfers within the Josephson-Majorana cycle are always correlated, regardless of the chosen parameters.
For this purpose, we study the short-time limit, $\Gamma t\ll1$ in a similar way as done in Ref.~\cite{stegmann2016short}.
We expand the generating function according to
\begin{align}
e^{{\cal{S}}(z,t)}&\approx 1+ \tr \left( {\cal L}\z\rho_\text{st}\right) t+\frac{1}{2}\tr \left( {\cal L}\z^2\rho_\text{st}\right) t^2  \nonumber \\
&= 1+ (z-1)P_1(t)+ (z-1)^2P_2(t).
\end{align}
To arrive at the second line, we identify the probabilities 
\begin{align}
P_1(t)&\approx  \tr({\cal J}\rho_\text{st})t, \\
P_2(t)&\approx \frac{1}{2} \tr({\cal J}^2\rho_\text{st})t^2,
\end{align}
where we used that $\tr({\cal L} \ldots )=0$ as well as ${\cal L}\rho_\text{st}=0$. 
Note that the probabilities fulfill $P_1(t)\propto t$ and $P_2(t)\propto t^2$ for a consistent perturbation expansion. 
A dynamical Lee-Yang zero analysis~\cite{stegmann2016short,kleinherbers2018revealing} reveals that whenever $4P_2>P_1^2$ is fulfilled, correlations are present in the statistics. Using Eq.~\eqref{eq:liouvillian_cycle}, we find for the Josephson-Majorana cycle
\begin{align}
\frac{4P_2}{P_1^2}=\frac{3 E_\text{J}^2+\Gamma^2+4 \omega^2}{2 E_\text{J}^2} \, .
\end{align}
This fraction is always larger than 1.
Hence, we have shown that correlations are indeed present for all parameters.

\begin{figure*}[t]
	\includegraphics[width=.95\textwidth]{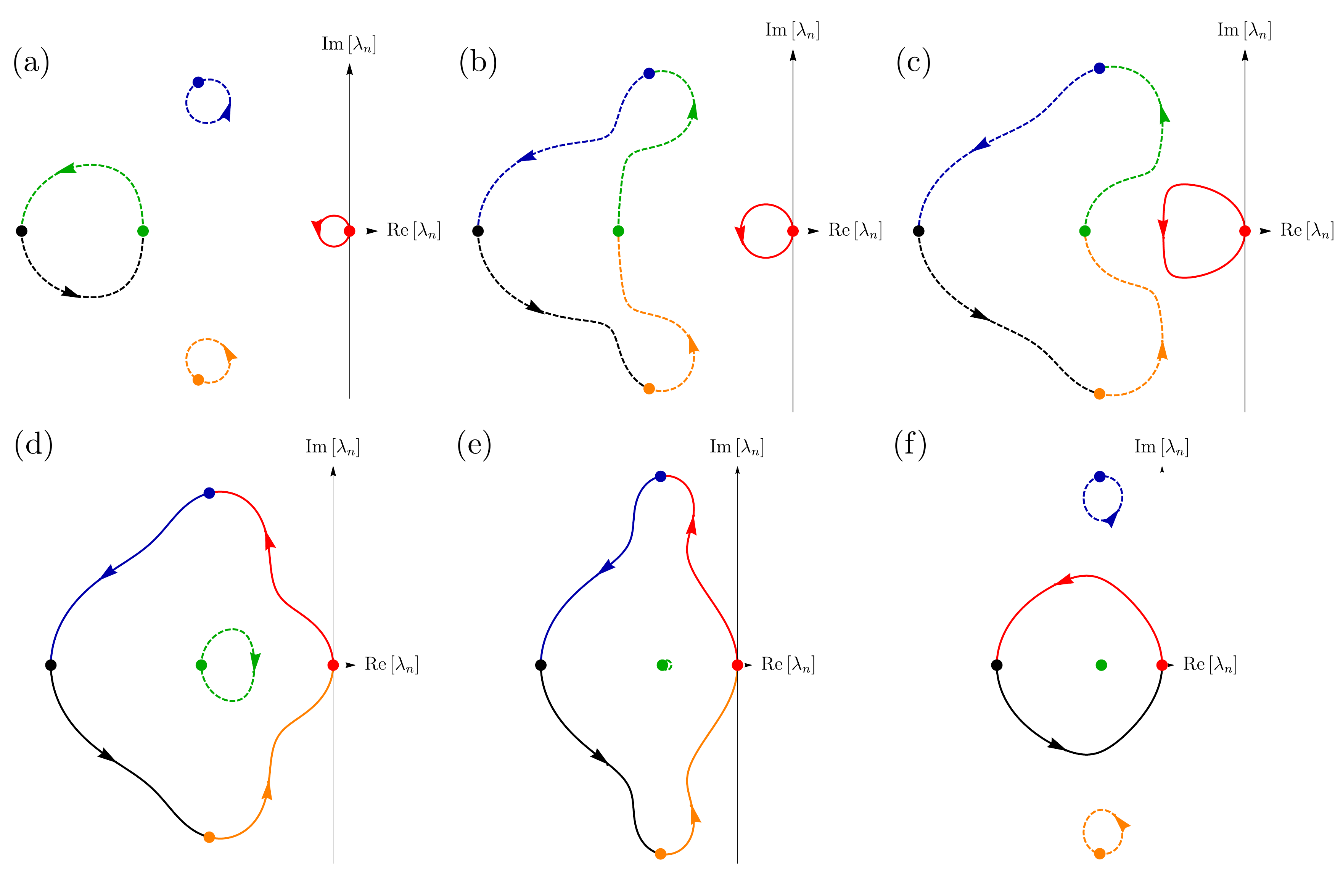}
	\caption{Topological transitions in the full counting statistics of electron transport indicated by the five nondegenerate eigenvalues $\lambda_{l}(z)$ evaluated on the unit circle $z=e^{i\chi}$ from $0<\chi<\pi$. The eigenvalues are colored red ($l=1$), blue ($l=2$), orange ($l=3$), green ($l=4$), and black ($l=5$). Solid lines indicate the primary loop that determines the counting statistics and dashed lines indicate the secondary loops. The colored dots show the eigenvalues at $\chi=0$. The parameters are (a) $E_\text{J}=0.3\Gamma$, (b) $E_\text{J}=0.4\Gamma$, (c) $E_\text{J}=0.53\Gamma$, (d) $E_\text{J}=0.6\Gamma$, (e) $E_\text{J}=1.2\Gamma$, and (f) $E_\text{J}=1.5\Gamma$. The charging energy is given by $E_\text{C}=1000\,\Gamma$.
}
\label{fig:topology}	
\end{figure*}

\subsection{Topological transition}
To better understand the changes in the counting statistics observed from Fig.~\ref{fig:facs}(a)-(c), we follow Ref.~\cite{riwar_2019} and study topological transitions of the full counting statistics in the long-time limit.
For $\Gamma t\gg 1$, only the eigenvalue $\lambda_\text{max}(z)$ of ${\cal L}\z$ with the largest real part (or, since the real parts are nonpositive, the smallest magnitude of the real part) determines the statistics
\begin{align}\label{eq:cumgen}
{\cal S}(z,t)\approx \lambda_\text{max}(z) t.
\end{align}
For the calculation of the factorial cumulants, only the derivatives of the cumulant-generating function ${\cal S}(z,t)$ with respect to $z$ at $z=1$ are needed.
In order to discuss topological transitions in the full counting statistics, however, we consider ${\cal S}(z,t)$ as a function of $z$ in the entire complex plane. 
Furthermore, we parametrize the counting variable $z$ via $z=e^{i\chi}$ by the counting field $\chi$ and restrict ourselves to real values of $\chi$, such that $z$ stays on the unit circle in the complex plane. 
We remark here in passing that performing derivatives of ${\cal S}(e^{i\chi},t)$ with respect to $\chi$ at $\chi=0$ generates ordinary cumulants instead of factorial ones.
Furthermore, the functional dependence of the cumulant-generating function on $\chi$ beyond $\chi=0$ can be used to distinguish different transport mechanisms contributing to the same order of a perturbation expansion in the coupling strength, such as cotunneling versus sequential tunneling with renormalized parameters~\cite{braggio_2006}.
The main motivation to parametrize $z$ by $\chi$, however, is that the spectrum of the Liouvillian ${\cal L}\z$, and therefore also $\cal S$, is \textit{periodic} as a function of $\chi$.
The topological transition which we want to discuss here is indicated by a change of the periodicity of the cumulant-generating function~\cite{riwar_2019,ren_2013}.

In Fig.~\ref{fig:topology}(a)-(f), the real and imaginary part of the full spectrum $\{ \lambda_l(e^{i\chi}) \}_{l=1,\ldots, 5}$ with $l=1,\ldots,5$, of the Liouvillian ${\cal L}\z$ is shown as a function of the counting field $\chi$. Starting at $\chi=0$, where the eigenvalues are indicated by colored dots, we indicate the change of the spectrum as the counting field is increased up to $\chi=\pi$ by directed lines. 
It is interesting to note that the entirety of the spectrum is periodic with periodicity $\pi$, which also can be easily checked analytically by using the characteristic polynomial $\det(\lambda-{\cal L}\z)\vert_{z=e^{i\chi}}$.
The periodicity of $\pi$ is associated with the joint transfer of two charges $2e$, in our case the normal and the anomalous tunneling that occur together within the Josephson-Majorana cycle.

The fact that the spectrum is periodic with $\pi$ does not necessarily mean that each eigenvalue has the same periodicity.
In contrast, it is possible that two or more eigenvalues are exchanged upon $\chi \rightarrow \chi+\pi$, which is referred to as braiding~\cite{ren_2013}. 
It is easy to identify such a scenario in Fig.~\ref{fig:topology}(a)-(f).
Braiding occurs whenever the shown loops contain more than one color, i.e., more than one eigenvalue.
For example, in Fig.~\ref{fig:topology}(e), there is a loop containing four eigenvalues.
The position of the four eigenvalues (red,blue,black,orange) permutates to (blue,black,orange,red) for $\chi =\pi$.
Only after a period of $4\pi$, the eigenvalues return to their initial position after completely winding around each other.
The increased periodicity to $4\pi$ could be associated with a fractional charge~\cite{riwar_2019,javed_2023} of $e/2$.

Now, we can analyze how the topology of the counting statistics changes from Fig.~\ref{fig:topology}(a)-(f) as the Josephson coupling $E_\text{J}$ is increased.
Due to Eq.~\eqref{eq:cumgen}, in the long-time limit all information is contained in the \textit{primary loop} (solid line) that passes through the origin. 
In Fig.~\ref{fig:topology}(a), for $E_\text{J}\ll\Gamma$, this primary loop contains only one eigenvalue that indeed returns to itself after a period of $\pi$. 
This is consistent with the $\chi$-dependence of the eigenvalue
\begin{align}\label{eq:poiss}
\lambda_\text{max}\approx \frac{E_\text{J}^2}{\Gamma^2+4\omega^2}\left(e^{2i\chi}-1\right)\Gamma \, ,
\end{align}
that we obtained by an explicit Taylor expansion of the spectrum in the Josephson coupling $E_\text{J}$. 
This is the cumulant-generating function of a perfect Poisson process of Cooper pairs, where only the first two factorial cumulants are nonzero, $C_{\text{F},1}= C_{\text{F},2}=\frac{2E_\text{J}^2}{\Gamma^2+4\omega^2}\Gamma t$, in accordance with Fig.~\ref{fig:facs}(a).

In Fig.~\ref{fig:topology}(b), $E_\text{J}$ is slightly increased.
While the topology of the secondary loops (dashed lines) have changed from three loops to one bigger loop containing four eigenvalues, the primary loop (solid line) remains qualitatively unchanged.
It still contains only one eigenvalue, and Eq.~\eqref{eq:poiss} is still a valid description of the statistics.

In Fig.~\ref{fig:topology}(c)-(d), the Josephson coupling is further increased and we get a topological transition when the primary and secondary loop touch each other.
After the transition, the primary loop contains four eigenvalues, while a new secondary loop appears in the interior of the primary one.
In Fig.~\ref{fig:topology}(e), merely the shape of the primary loop changes. In addition, the size of the detached secondary loop shrinks almost to a point.

Finally, in Fig.~\ref{fig:topology}(f), two secondary loops detach from the primary one and we are left with two eigenvalues with a $2\pi$ periodicity each. 
In this case, the cumulant-generating function can asymptotically (for $E_\text{J}\gg\Gamma$) be described by a square root 
\begin{align}
\lambda_\text{max}\approx \frac{\Gamma}{4} \left(-3+\sqrt{1+8e^{2i\chi}}\right).
\end{align}
This cumulant-generating function is consistent with the switching dynamics between even- and odd-parity states as described by Eq.~\eqref{eq:switching}.

In summary, we find upon increasing the Josephson coupling two topological transitions of the full counting statistics.
First, the periodicity of the relevant eigenvalue changes from $\pi$ to $4 \pi$ and then, later, from $4\pi$ to $2\pi$.

Finally, we remark that the notion of topology in this paper appears twice in mutually unrelated ways. 
On the one hand, the Majorana single-charge transistor relies on the formation of topologically protected Majorana bound states.
On the other hand, the full counting statistics features topological transitions related to the periodicity of the cumulant-generating function as function of the counting field.
To avoid confusion, we mention that topological transitions in the full counting statistics are also present in topologically-trivial systems such as single and double quantum dots~\cite{riwar_2019}.

%%%%%%%%%%%%%%%%%%%%%%%%%%%%%%%%%%%%%%%%%%%%%%%%%%%%%%%%%%%%%%%%%%%%%%%%%%%%%%%%%%%%%%%%%%%%%%%%%%%%
%%%%%%%%%%%%%%%%%%%%%%%%%%%%%%%%%%%%%%%%%%%%%%%%%%%%%%%%%%%%%%%%%%%%%%%%%%%%%%%%%%%%%%%%%%%%%%%%%%%%
\section{\label{sec:conclusions} Conclusions}
We studied the electron transport through a Majorana single-charge transistor coupled to a normal and a superconducting lead. 
At low bias voltages, the dominant transport mechanism is the
Josephson-Majorana cycle, which may, depending on the parameters, be suppressed via a normal or coherent Coulomb blockade.
To address the correlated nature of single-charge transfers within the Josephson-Majorana cycle, we calculated the waiting-time distribution and the full counting statistics in terms of factorial cumulants. Using a sign criterion for factorial cumulants to indicate correlations, we could prove that the electron transfer in the Josephson-Majorana cycle is highly correlated. 
This may serve as a motivation trying to experimentally implement a charge detector in order to monitor full time traces of the individual charge transfers. 
As we have shown, this opens the possibility to assess the nature of the underlying transport process in a way that would not be possible by measuring the average current only.

Moreover, by means of the spectrum of the Liouvillian and the winding of its eigenvalues as a function of the counting field, we identify topological transitions in the full counting statistics. 
We found that the stochastic process changes from a perfect Poisson process of two charges at small Josephson coupling to a dynamic switching between even- and odd-parity states at large coupling.
The study of topological transitions in the full counting statistics of nanoscale devices is a rather new research topic. 
The Majorana single-charge transistor provides a nice and non-trivial model system for studying such transitions.

\begin{acknowledgments}
This work was supported by the Deutsche Forschungsgemeinschaft (DFG, German  Research Foundation) under Project-ID 278162697 -- SFB 1242.
\end{acknowledgments}

\appendix
\section{Generalized master equation}\label{details}
The tunneling-induced dynamics is described by the superoperator ${\cal W}$ in the Liouville equation Eq.~\eqref{eq:mastereq} and is defined via
\begin{align}\label{eq:tunneling}
{\cal W} \rho=\sum_{\Delta E,\Delta E^\prime,\pm}&\Gamma_\pm(\Delta E)\Big(\hat{d}_{\pm, \Delta E}^{\phantom{dagger}} \rho\, \hat{d}_{\pm,\Delta E^\prime}^\dagger \nonumber \\
&-\hat{d}_{\pm, \Delta E^\prime}^\dagger \hat{d}_{\pm, \Delta E}^{\phantom{dagger}} \rho \Big)+\text{h.c.},
\end{align}
where the rate $\Gamma_\pm(\Delta E)$ is given by Eq.~\eqref{eq:rate} and we introduced the energy-resolved excitations 
\begin{align}
\hat{d}_{+,\Delta E}&=\sum_{\chi,\chi^\prime}\delta_{\Delta E,E_{\chi}-E_{\chi^\prime}}\mel{\chi}{\hat{d}^\dagger {-} e^{i\hat{\varphi}} \hat{d}}{\chi^\prime} \dyad{\chi}{\chi^\prime},\\
\hat{d}_{-,\Delta E}&=\sum_{\chi,\chi^\prime}\delta_{\Delta E,E_{\chi^\prime}-E_{\chi}}\mel{\chi}{\hat{d} {-} e^{-i\hat{\varphi}} \hat{d}^\dagger}{\chi^\prime} \dyad{\chi}{\chi^\prime},
\end{align}
where $\ket{\chi}$ and $\ket{\chi^\prime}$ indicate the eigenstates of $H_\text{C}+H_\text{J}$. 
Here, the first and second line describe processes where effectively a single charge enters ($+$) and leaves ($-$) the TSI, respectively. The corresponding excitation energies are labeled by $\Delta E$.
In the diagrammatic picture, terms of the form $\sim \hat{d}_{\pm, \Delta E^\prime}^\dagger \hat{d}_{\pm, \Delta E}^{\phantom{dagger}} \rho$ and $\sim\rho\, \hat{d}_{\pm, \Delta E^\prime}^\dagger \hat{d}_{\pm, \Delta E}^{\phantom{dagger}} $ of Eq.~\eqref{eq:tunneling} originate from diagrams acting solely on one branch of the Keldysh contour, while terms of the form $\sim \hat{d}_{\pm, \Delta E}^{\phantom{dagger}} \rho\, \hat{d}_{\pm,\Delta E^\prime}^\dagger$ originate from diagrams connecting both branches~\cite{kleinherbers_2020}.
Now, we can also define the jump operators 
\begin{align}
{\cal J}_\pm \rho=\sum_{\Delta E,\Delta E^\prime}\left[\Gamma_\pm(\Delta E)+\Gamma^*_\pm(\Delta E^\prime)\right]\hat{d}_{\pm, \Delta E}^{\phantom{dagger}} \rho\, \hat{d}_{\pm,\Delta E^\prime}^\dagger,
\end{align}
which can be used to calculate the net current using Eq.~\eqref{eq:current}.

\section{Secular approximation}\label{app:rate}
For $E_\text{J}\gg\Gamma$, the dynamics of the Josephson-Majorana cycle (around $n_V=1$) is well described by the Liouvillian ${\cal L}^\text{S}_{z}$ in the secular approximation.  The latter can be determined from the full Liouvillian ${\cal L}_{z}={\cal L}_{z,z}$ of Eq.~\eqref{eq:liouvillian_cycle} via an average in the interaction picture~\cite{kleinherbers_2020}
\begin{align}
{\cal L}^\text{S}_{z}=\lim_{T\rightarrow\infty}\frac{1}{2T}\int\limits_{-T}^{T} \mathrm{d}t \,e^{-{\cal L}_0t}{\cal L}_{z}\,e^{{\cal L}_0t},
\end{align}
where we identified the superoperator ${\cal L}_0=-\frac{i}{\hbar} \left[ H_\text{C}+H_\text{J}, \ldots \right]$ describing the von Neumann part of the time evolution. Switching to the eigenbasis $\rho_{z}=(\rho_{\Psi_1},\rho_{\Psi_3},\rho_{\Psi_4},\rho_{\Psi_4}^{\Psi_3},\rho_{\Psi_3}^{\Psi_4})_{z}$ (where $\Psi_1,\Psi_3$, and $\Psi_4$ indicate the three relevant eigenstates given in Eq.~\eqref{eq:eigenstates} for $\theta=0$), we obtain
\begin{align}
{\cal L}^\text{S}_z=\begin{pmatrix}
 -\Gamma  &  z \frac{\Gamma}{2} & z \frac{\Gamma}{2}   & 0 & 0 \\
z\frac{\Gamma}{2}  & -\frac{\Gamma}{2} & 0 & 0 & 0 \\
z\frac{\Gamma}{2}  & 0 &  -\frac{\Gamma}{2}  & 0 & 0 \\
 0 & 0 & 0& -\frac{\Gamma }{2}+i E_\text{J} & 0 \\
 0 & 0 & 0 & 0 & -\frac{\Gamma }{2} -i E_\text{J}\\
\end{pmatrix},
\end{align}
so that the coherences completely decouple from the populations. Defining further the probabilities $p_\text{odd}=\rho_{\Psi_1}$ and $p_\text{even}=\rho_{\Psi_3}+\rho_{\Psi_4}$ to find an odd- and an even-parity state, we arrive at the simple rate equation of Eq.~\eqref{eq:switching}.

%%%%%%%%%%%%%%%%%%%%%%%%%%%%%%%%%%%%%%%%%%%%%%%%%%%%%%%%%%%%%%%%%%%%%%%%%%%%%%%%%%%%%%%%%%%%%%%%%%%%
%%%%%%%%%%%%%%%%%%%%%%%%%%%%%%%%%%%%%%%%%%%%%%%%%%%%%%%%%%%%%%%%%%%%%%%%%%%%%%%%%%%%%%%%%%%%%%%%%%%%
\bibliography{bibliography}

%%%%%%%%%%%%%%%%%%%%%%%%%%%%%%%%%%%%%%%%%%%%%%%%%%%%%%%%%%%%%%%%%%%%%%%%%%%%%%%%%%%%%%%%%%%%%%%%%%%%
%%%%%%%%%%%%%%%%%%%%%%%%%%%%%%%%%%%%%%%%%%%%%%%%%%%%%%%%%%%%%%%%%%%%%%%%%%%%%%%%%%%%%%%%%%%%%%%%%%%%
\end{document}